\documentclass[aps,twocolumn,,superscriptaddress,showpacs,showkeys,amsmath,amssymb,floatfix]{revtex4}
\usepackage{graphicx}% Include figure files
\usepackage{epsfig}
\usepackage{dcolumn}% Align table columns on decimal point
\usepackage{bm}% bold math
\usepackage{amssymb}
\usepackage{dsfont}
\usepackage{amsmath}
\usepackage{subfigure}
\newcommand{\be}{\begin{equation}}
\newcommand{\ee}{\end{equation}}
\newcommand{\bea}{\begin{eqnarray}}
\newcommand{\eea}{\end{eqnarray}}

\newcommand{\hD}{{\hat D}}

\newcommand{\tC}{{\tilde C}}

\newcommand{\tzeta}{{\tilde \zeta}}

\newcommand{\vA}{\vec A}
\newcommand{\vF}{\vec F}

\newcommand{\vX}{{\vec X}}
\newcommand{\pro}{\partial}
\newcommand{\der}{\partial}

\newcommand{\bphi}{{\mathbf \phi}}

\newcommand{\ba}{\begin{array}}
\newcommand{\ea}{\end{array}}

\newcommand{\nn}{\nonumber}

\newcommand{\uast}{\stackrel{\ast}{u}}
\begin{document}
\title{Non-monopole magnetic solutions in the Weinberg-Salam model}
\bigskip
\author{A. S. Bakry}
\affiliation{Institute of Modern Physics, Chinese Academy of Sciences,
Lanzhou 730000, China}
\author{D. G. Pak}
\affiliation{Institute of Modern Physics, Chinese Academy of Sciences,
Lanzhou 730000, China}
\affiliation{Laboratory of Few Nucleon Systems,
Institute for Nuclear Physics, Ulugbek, 100214, Uzbekistan}
\author{P. M. Zhang}
\affiliation{Institute of Modern Physics, Chinese Academy of Sciences,
Lanzhou 730000, China}
\affiliation{State Key Laboratory of Theoretical
Physics, Institute of Theoretical Physics,
Chinese Academy of Sciences, Beijing 100190, China}
\author{L. P. Zou}
\affiliation{Institute of Modern Physics, Chinese Academy of Sciences,
Lanzhou 730000, China}

\begin{abstract}
The structure of finite energy non-monopole solutions
with azimuthal magnetic flux of topological origin
is studied in the pure bosonic sector of the Weinberg-Salam model.
Applying a variational method we have found simple magnetic field
configurations which minimize the energy functional
and possess energies of order 1 TeV.
Such configurations correspond to composite bound states
of $W,Z$ and Higgs bosons with essentially less energy
in comparison to monopole like particles supposed to be found at LHC.
\end{abstract}
%\vspace{0.3cm}
\maketitle
%\vspace{2mm}

\section{Introduction}

The search for stable massive particles in collider experiments
has become of great importance \cite{fairb}. Electroweak monopoles
represent one of the possible candidates for new fundamental particles
expected to be found at LHC \cite{milton, CERN,IJMPA}.
Known singular monopole solutions like the Dirac monopole
\cite{dirac} and its generalizations \cite{wuyang,nambu77,chomaison}
have infinite energy, so the main characteristic of particles, the mass,
represents a free parameter which can not be deduced from the
theory. Moreover, it has been proved that for a wide class of axially-symmetric
magnetic fields any finite energy monopole solution must have a totally
screened magnetic charge \cite{ws1}.
%Moreover, a consistent quantum description
%of such monopoles requires introduction of a respective monopole
%charged matter field (described by Dirac fermion or complex scalar field) and
%axial magnetic current which will imply quantum anomalies in the theory.
%This implies a cardinal extension of the standard model so it makes
%doubts on search of such solely existing monopoles.
%Composite monopoles like 't Hooft-Polyakov one \cite{thooft,polyakov74}
%give an alternative construction of monopole like solutions
%without need to introduce a new fundamental particle.
Another candidate for a monopole-like particle is the
so-called "monopolium" which represents
a monopole-antimonopole bound state.
A known unstable sphaleron solution \cite{DHN,manton83,klinkmanton84}
might provide a theoretical basis for the existence of
such a bound state due to interpretation of the sphaleron
as a monopole-antimonopole pair \cite{hind94, ws1}.

In the present paper we consider properties
of a possible magnetic solution with azimuthal magnetic flux.
We have found that energy of such a solution can be much less than
energy 7-8 TeV for monopole-like solutions (Cho-Maisson monopole,
sphaleron).
Since the magnetic field configuration with the
azimuthal magnetic flux belongs to a general
axially-symmetric class, to find a strict solution
one must solve a full set of highly non-linear
partial differential equations
of motion which is currently a difficult numeric problem.
So we apply a variational energy minimization procedure
to obtain qualitative description of the energy density profile
of the magnetic solution.
We have found two magnetic field configurations minimizing the energy functional
with energies $1.36$ TeV and $0.98$ TeV depending on the choice
of boundary conditions for the gauge fields and Higgs boson.
Both configurations have similar qualitative structure with energy
density maximums located along two parallel rings
with centers on the $Z$-axis.
We conjecture the existence of such magnetic solutions
which would correspond to magnetic bound states of $W,Z$
and Higgs bosons with energies of order 1 TeV.

\section{Axially-symmetric ansatz}

Let us start with the Lagrangian of the Weinberg-Salam model
describing the pure bosonic sector
(we use Greek letters for space-time indices, $\mu,\nu=0,1,2,3$,
and Latin letters for space vector and $SU(2)$ internal indices, $a,b=1,2,3$)
\bea
{\cal L} =
-\dfrac{1}{4} {\vec F}_{\mu\nu}^2-\dfrac{1}{4}G_{\mu\nu}^2
  - |D_\mu \bphi|^2-\dfrac{\lambda}{2} ({\bphi}^\dagger {\bphi} -
 \dfrac{v^2}{2})^2,
%&& {F}_{\mu\nu}^a=\pro_\mu A_\nu^a-\pro_\nu A_\mu^a +g \epsilon^{abc} A_\mu^b A_\nu^c, \nn \\
%&&G_{\mu\nu}=\pro_\mu B_\nu - \pro_\nu B_\mu, \nn \\
%&&D_\mu=\pro_\mu -\dfrac{ig}{2} \vec \sigma \cdot \vec A_\mu -i \dfrac{g'}{2} B_\mu,
\label{LagrWS}
\eea
where $\vec F_{\mu\nu}$ and $ G_{\mu\nu}$ are gauge field strengths,
$\bphi$ is the Higgs complex scalar doublet and
$\vA_\mu$ and $B_\mu$ are the gauge potentials
of the electroweak gauge group $SU(2)\times U_Y(1)$.
The Higgs field can be parameterized in terms of a scalar field $\rho(x)$,
a unit complex $SU(2)$ doublet $\tzeta$ and $U_Y(1)$ field variable $\omega(x)$
in the exponential factor as follows
\bea
&&\bphi= \dfrac{1}{\sqrt 2} \rho \tzeta e^{i\omega(x)},~~~~\tzeta^+ \tzeta=1. \label{higgspar}
\eea
%It has been observed \cite{chomaison} a crucial difference in topology
%of the Yang-Mills-Higgs theory with a complex Higgs doublet and the
%Weinberg-Salam model.
In the Weinberg-Salam model due to the presence of the local
$U_Y(1)$ symmetry one can factorize $U_Y(1)$ degree of freedom by imposing
a gauge $\omega(x)=0$. One should stress that
such a possibility is absent in a pure Yang-Mills-Higgs model.
This implies that the Higgs complex field $\tzeta$ describes
a two-dimensional sphere and admits non-trivial homotopy groups
$\pi_{2,3}(S^2)$ which provide necessary conditions
for finite energy monopole solutions at least at space infinity
\cite{chomaison}. Notice, the gauge fixing condition
$\omega=0$ is consistent with the unitary gauge for the Higgs field.
Moreover, the Weinberg-Salam model in such a fixed gauge is equivalent
to the original theory at classical and quantum level within perturbation theory
due to absence of any ghosts. On the other hand, the un-fixed field degree of freedom $\omega$
implies that the homotopy group $\pi_{2,3}(S^2)$ represents
in a fact a relative homotopy, so the topological structures of the
theory with fixed and un-fixed $U_Y(1)$ symmetry are different.
This shows clearly the lack of deep understanding the origin of
the spontaneous symmetry breaking in the electroweak theory.

In the previous paper we have shown that magnetic field screening
effect leads to non-existence of solutions representing
a system of single monopoles and antimonopoles \cite{ws1}.
So we will consider non-monopole solutions which
possess non-trivial magnetic fluxes of topological origin.
We apply a gauge invariant decomposition for the $SU(2)$ gauge potential
\cite{choprd80,duan} which allows to trace the topological structure
and features of the interaction of the Higgs and gauge bosons.
In particular, we will show that interaction structure of the
$W,Z$ and Higgs bosons implies essential total energy decrease
for magnetic solutions.

First we construct a unit triplet vector field $\hat m$ explicitly
through the Higgs field
\bea
\hat m^a&=&\dfrac{1}{|\phi|^2} \phi^+ {\vec \sigma}^a \phi=\tzeta^+ {\vec \sigma}^a \tzeta,
%\hat \phi&=& \dfrac{\phi}{|\phi|},
\label{hatm}
\eea
where $\vec \sigma^a$ are Pauli matrices.
With this one can perform gauge invariant decomposition
of the $SU(2)$ gauge potential into Abelian and off-diagonal parts \cite{choprd80,duan}
\bea
\vA_\mu&=&A_\mu \hat m+\vec C_\mu +\vX_\mu,  \nn \\
\vec C_\mu&=&-\dfrac{1}{g} \hat m \times \pro_\mu \hat m,~~~~~
(\vec X_\mu \cdot \hat m) =0 , \label{chodec}
\eea
where the vector potential $\vec C_m$ is made of the Higgs vector field $\hat m$,
and $\vX_\mu$ represents two off-diagonal components of the gauge potential $\vA_\mu$.
Respectively, one has the following Abelian decomposition
for the full $SU(2)$ gauge field strength
\bea
\vF_{\mu\nu}&=&(F_{\mu\nu}+H_{\mu\nu}) \hat m \nn +\\
 &&\hD_\mu \vX_\nu-\hD_\nu \vX_\mu+g \vX_\mu \times \vX_\nu, \nn \\
 F_{\mu\nu}&=& \pro_\mu A_\nu-\pro_\nu A_\mu, \nn \\
H_{\mu\nu}&=&\dfrac{1}{g}\epsilon^{abc}\hat m^a \der_\mu \hat m^b \der_\nu \hat m^c=
               \pro_\mu \tC_\nu-\pro_\nu \tC_\mu, \label{cdual}
\eea
where $\tC_\mu=2i\tzeta^+\pro_\mu \tzeta$ is a dual magnetic potential,
and, $\hat D_\mu=\pro_\mu+A_\mu \hat m+\vec C_\mu$ is a restricted $SU(2)$ covariant derivative \cite{choprd80}.
One can define an $SU(2)$ gauge invariant Abelian gauge potential ${\cal A}_\mu$
and a respective Abelian gauge field strength ${\cal F}_{\mu\nu}$ as follows \cite{ws1}
\bea
{\cal A}_\mu&=&A_\mu+\tilde C_\mu,~~~ \nn \\
A_\mu&=& \vec A_\mu \cdot \hat m, \nn \\
{\cal F}_{\mu\nu}&=&F_{\mu\nu}+H_{\mu\nu}, \label{calA}
\eea
where $F_{\mu\nu},~ H_{\mu\nu}$ are Maxwell type field strengths.
One should stress the importance of the introduced
above gauge invariant quantities ${\cal A}_\mu$ and
${\cal F}_{\mu\nu}$ representing a composite combination
of $SU(2)$ gauge bosons and Higgs. Due to additive structure
of the field strength it becomes clear that contributions
of the Higgs and $W,Z$ bosons can be mutually canceled
under appropriate conditions. This provides the origin of
drastic energy decrease in solutions as we will show below.

With this one can introduce the magnetic charge and a generalized Chern-Simons number
in invariant manner with respect to local $SU(2)$ gauge transformation
\bea
{\cal Q}_m&=&\dfrac{1}{A({\cal S})}\int_{{\cal S}^2} {\cal F}_{ij} \cdot d \sigma^{ij}, \nn \\
{\cal Q}_{CS}&=& \dfrac{1}{32 \pi^2}\int d^3x \epsilon^{ijk} {\cal A}_i {\cal F}_{jk}, \label{gimonHopfch}
\eea
where $A({\cal S})$ is the area of a closed two-dimensional surface ${\cal S}^2$.
A unique definition of the electromagnetic vector potential
and the neutral gauge $Z$-boson is given by the following expressions
\bea
&&A^{em}_\mu=\cos \theta_W  B_\mu + \sin \theta_W {\cal A}_\mu, \nn \\
&&Z_\mu=-\sin \theta_W B_\mu + \cos \theta_W {\cal A}_\mu. \label{Aem}
\eea

A crucial feature of the gauge invariant decomposition (\ref{chodec},\ref{cdual}) is
that Abelian projection  of the gauge
field strength $\vec F_{\mu\nu}$ onto the direction along the vector $\hat m$
includes the Abelian field $F_{\mu\nu}$
and the magnetic field $H_{\mu\nu}$ made of the Higgs field
in additive form. This implies the following expression
for the Yang-Mills part of the Weinberg-Salam Lagrangian (\ref{LagrWS})
\bea
{\cal L}_{SU(2)}&=&\dfrac{1}{4}(F_{\mu\nu}+H_{\mu\nu}+g \hat m\cdot [\vX_\mu \times \vX_\nu])^2+\nn  \\
 && \dfrac{1}{4}(\hD_\mu \vX_\nu-\hD_\nu \vX_\mu)^2. \label{Lagrred}
 \eea
%Now it becomes clear, the Abelian magnetic fields $F_{\mu\nu}$
%and $H_{\mu\nu}$ may cancel each other
%in local space regions decreasing the total energy of the field configuration.
Obviously, the above would entail  that
the Abelian magnetic fields $F_{\mu\nu}$ and $H_{\mu\nu}$
 can cancel partially each other in local space regions, thus
decreasing the total energy of the field configuration.
%Since any physical solution represents a local minimum of the
%energy functional such mutual cancellation of the gauge and Higgs bosons
%will provide magnetic field screening phenomenon.

For our further purpose it is convenient
to express the Higgs field $\hat m$ in terms
of one complex scalar function
using a standard stereographic projection
\bea
\hat m &=& \dfrac{1}{1+u \uast}
 \left (\ba{c}
  u+\uast\\
  -i (u-\uast)\\
 u \uast-1\\
            \ea
           \right ), \nn \\
u &=&\dfrac{\tzeta_1}{\tzeta_2}.\label{nstereo}
\eea
We will apply a most general ansatz
for static axially-symmetric magnetic fields \cite{ws1}
which includes axially-symmetric gauge potentials
$A_\mu(r,\theta),~\vec X_\mu(r,\theta)$ and the
Higgs scalar $\rho(r,\theta)$
in spherical coordinates $(r,~0\leq\theta\leq\pi, ~0\leq\varphi\leq2 \pi)$.
For the Higgs vector $\hat m$
we adopt an ansatz with two independent field variables
$f(r,\theta),~ Q(r,\theta)$
\bea
u(r,\theta,\varphi)&=& e^{-i m \varphi} \big (\cot(\dfrac{n\theta}{2}) f(r,\theta)+
i \csc (\dfrac{n\theta}{2})Q(r,\theta)\big), \nn \\
&&\label{ansfQ}
\eea
where integer numbers $(m,~n)$ determine
the topological monopole and
Hopf charges of the vector field $\hat m$.
%\bea
%Q_m&=&\dfrac{1}{A(S)}\int_{S^2} H_{ij} \cdot d \sigma^{ij}, \nn \\
%Q_H&=& \dfrac{1}{32 \pi^2}\int d^3x \epsilon^{ijk} \tC_i H_{jk} , \label{monHopfch}
%\eea
%where $A(S)$ is the surface area of a sphere $S^2$.

For convenience, the energy functional can be expressed in
terms of dimensionless variables
$ r \rightarrow r m_W$,
 $k \equiv  2 \sqrt \lambda /g $,
$\vA_\mu\rightarrow \vA_\mu g/m_W$, $\tC_\mu\rightarrow \tC_\mu g/m_W$,
 $B_\mu\rightarrow B_\mu g'/m_W$, $\rho \rightarrow \rho/v $.
 The energy functional of the Weinberg-Salam model
for static magnetic fields would then read
%With this the energy functional of the Weinberg-Salam model
%for a case of static magnetic fields reads
\bea
E &=& \dfrac{m_W}{g^2}  \int d^3 x \Big [\dfrac{1}{4} {\vec F_{mn}}^2
                     +\dfrac{1}{4}\kappa {\vec G_{mn}}^2 + \nn \\
           &&  2 |D_m \bphi|^2 + \dfrac{k^2}{2} (\rho^2-1)^2 \Big ], \label{energy}
\eea
where $\kappa = \dfrac{g^2}{g'^2}=3.487$,
$k^2=2.418$ and
$m_W =80.387 $ Gev , $m_H=125 $ Gev, $\, \sin^2\theta_W=0.22286$.
The numeric value of the
mass factor in front of the integral is $\dfrac{m_W}{g^2}=195$ Gev.
%One can define a natural basis frame in the internal space of group $SU(2)$

\section{A simple $CP^1$ model}

Before constructing possible magnetic solutions with an
azimuthal magnetic flux in the Weinberg-Salam model,
let us first overview a similar magnetic solution
in the $CP^1$ model obtained by reduction of the Weinberg-Salam Lagrangian
\cite{ws1}.
Using Abelian decomposition (\ref{chodec}) we set $A_\mu=\vX_\mu=0,~$
and $\rho=1$. With this we obtain the following energy functional
from (\ref{energy})
\bea
E^{CP^1}&=&\dfrac{m_W}{g^2}\int d^3 x \Big [\dfrac{1}{4}
H_{mn}^2+\dfrac{1}{2}\vec C_m^2 \Big ]. \label{CP1en}
\eea
The energy functional $E^{CP^1}$ corresponds to
a modified Skyrme model where some exact
finite energy solutions have been obtained \cite{ferreira13}.
The existence of finite energy stable solutions
in the model determined by the energy functional (\ref{CP1en})
is provided by Derrick's theorem \cite{derrick}. Simple
scaling arguments imply that energy contributions from
 two terms in (\ref{CP1en}) are equalled. We make a further
simplification of the model by imposing a constraint $f(r,\theta)=1$.
Changing variable, $Q(r,\theta)=\cot(\dfrac{S(r,\theta)}{2})$,
one can write down the equation of motion for $S(r,\theta)$
as follows
%\begin{widetext}
\bea
%&&(r^2 \cos^2\theta+4\sin^2\theta \sin^2\dfrac{S}{2})S_{rr}+
%(\dfrac{1}{r^2} \sin^2\theta\sin^2S +
%\cos^2\theta)S_{\theta\theta}+\dfrac{1}{r^2} \sin^2\theta\sin S\big(\cos S S^2_\theta
%-4 \sin^2\dfrac{S}{2}\nn+ \\
%&&3 \cot\theta\sin S S_\theta \big )
%+ \dfrac{1}{4 \sin\theta} (\cos\theta+3\cos(3\theta)) S_\theta-
%2 \cos^2\theta(\sin S-r S_r)+\sin^2\theta \sin S S_r =0. \label{eqS}
&&(r^2 \cos^2\theta+4\sin^2\theta \sin^2\dfrac{S}{2})S_{rr}+
(\dfrac{1}{r^2} \sin^2\theta\sin^2S + \nn \\
&& \cos^2\theta)S_{\theta\theta}+\dfrac{1}{r^2} \sin^2\theta\sin S\big(\cos S S^2_\theta
-4 \sin^2\dfrac{S}{2}\nn+ \\
&&3 \cot\theta\sin S S_\theta \big )
+ \dfrac{1}{4 \sin\theta} (\cos\theta+3\cos(3\theta)) S_\theta- \nn \\
&&2 \cos^2\theta(\sin S-r S_r)+\sin^2\theta \sin S S_r =0. \label{eqS}
\eea
%\end{widetext}
Using finite energy condition one can find
appropriate boundary conditions
\bea
S(0,\theta)&=&0, ~~~~~S(\infty, \theta)=2 \pi. \label{boundflux}
\eea
Note that only multiple of $2 \pi$ values are allowed
for boundary conditions of $S(r,\theta)$.
The chosen boundary conditions (\ref{boundflux}) provide a
topological azimuthal magnetic flux $2 \pi$ through the half
plane $P:\{y=0,x\geq 0\}$.
We apply the numeric package COMSOL Multiphysics 3.5
to solve the partial differential equation (\ref{eqS}).
The solution for the function $S(r,\theta)$ is presented in Fig. 1.
%Ap13CP1-3extremF.mph, Ap12CP1-1.nb
\begin{figure}[htp]
\centering
\includegraphics[width=65mm,height=45mm]{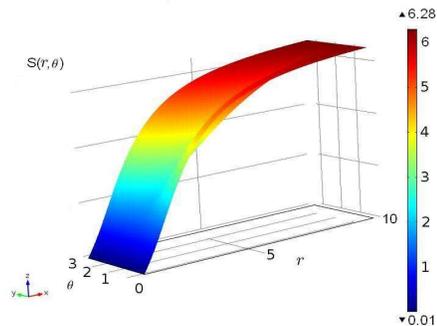}
\caption[fig1]{A three-dimensional plot for the solution $S(r,\theta)$ in the $CP^1$ model.}\label{Fig1}
\end{figure}
\begin{figure}[htp]
\centering
\includegraphics[width=65mm,height=45mm]{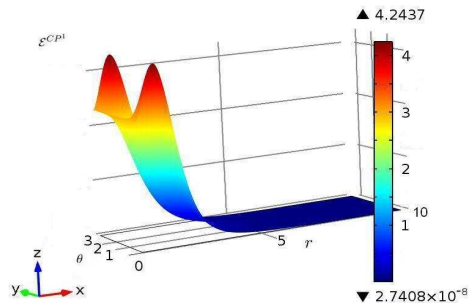}
\caption[fig2]{Energy density surface.}\label{Fig2}
\end{figure}

%\begin{figure}[htp]
%\centering
%\subfigure[~]{\includegraphics[width=35mm,height=35mm]{cp1-3.jpg}}
%\hfill
%\subfigure[~]{\includegraphics[width=40mm,height=40mm]{cp1-5c.jpg}}
%\caption[fig2]{(a) Function $S$; (b) Energy density}\label{Fig2}
%\end{figure}
The energy density ${\cal E}^{CP^1}$ corresponding to the solution
has two local maximums located on the $Z$-axis, Fig. 2.
The energy is $18$ TeV, and
the energy contribution from the first term in (\ref{CP1en})
is 51 \% of the total energy
in agreement with the analytic estimate based on Derrick's theorem.

Existence of the exact numeric solution with the azimuthal
magnetic flux in the Higgs subsector of the Weinberg-Salam model
provides a strong indication to existence of a similar solution in the full
theory where the Higgs boson is dressed by
gauge $W,Z$ bosons. From ordinary considerations the energy of such a solution
is expected to be of the same order.
It is surprising, a careful numeric analysis presented below
leads to energy estimate by one order less than $18$ TeV.

%\vspace*{3mm}
\section{Magnetic solution with azimuthal magnetic flux}

Let us consider possible magnetic solutions with a topological magnetic flux
around the $Z$-axis in the pure bosonic sector of the Weinberg-Salam model.
We apply a variational method to find field configurations which
minimize the energy functional (\ref{energy}) with boundary conditions
found from exact local solutions near the origin and space infinity.
For numeric purpose we change the variable
\bea
%f(r,\theta)=\dfrac{2}{F(r,\theta)}-1,~~~
Q(r,\theta)=\frac{1}{G(r,\theta)}-1,
\eea
and apply the following ansatz for static $SU(2)$ gauge potential
in the temporal gauge $\vec A_0=0$
\bea
\vec A_m&=& A_m \hat m-\hat m \times \pro_m \hat m \, W(r,\theta).
\eea
With additional constraints $A_r=B_r=0$ one has eight
trial functions $f,G,W,A_\theta, A_\varphi,B_\theta, B_\varphi,\rho$
dependent on two spherical coordinates $(r,\theta)$.

%Notice, in the known sphaleron solution the parametrization of the Higgs
%field is not most general due to vanishing function $f(r,\theta)$ in
%the ansatz (\ref{ansfQ}).
Note that, in \cite{ws1} we have studied a magnetic field
configuration with azimuthal magnetic flux which minimizes
the energy functional of Yang-Mills-Higgs theory.
% with a complex Higgs doublet.
The energy estimate $4.3 $ TeV for such a field configuration has been obtained
in a special case when a constrained ansatz for the Higgs vector field
$\hat m$ is applied, $f(r,\theta)=1$.
All known solutions in the Yang-Mills-Higgs theory and
in the Weinberg-Salam model (sphaleron, Cho-Maison monopole, monopole-antimonopole etc.)
are studied within the Dashen-Hasslacher-Neveu (DHN) ansatz \cite{DHN, manton83}
which contains the same constraint $f(r,\theta)=1$.
 It is worth noting that the function $f(r,\theta)$ represents an
independent degree of freedom of the Higgs boson and is
supposed to be determined dynamically by equations of motion.
So we conjecture that in general the field variable $f(r,\theta)$
should be a non-trivial function. We will consider two types of boundary conditions
which correspond to two different local solutions near the origin $r=0$.

%\vspace*{2mm}
\subsection{Type I boundary conditions}
%\vspace*{2mm}

Full equations of motion of the Weinberg-Salam model
represent highly non-linear partial differential equations for which
the boundary value problem admits regular solutions not for arbitrary boundary conditions.
To find proper boundary conditions one should solve equations of motion
near boundaries and check consistency with finite energy conditions.
We will find local solutions to the equations of motion of the Weinberg-Salam model
in the vicinity of the origin $r=0$ and in the
asymptotic infinity $r=\infty$ region.
Substituting Taylor series expansions for the trial functions
$f,G,W,A_\theta, A_\varphi,B_\theta, B_\varphi,\rho$
into all equations of motion one obtains
the following local solution near the origin
in lowest order approximation
%{\scriptsize {\bf A7M10M8Jul13G0=0CHECKSOLcaseF=0.nb}}
\bea
f(r,\theta)&=&1-2 C_1 r^2 (5 \cos^2\theta-1), \nn  \\
G(r,\theta)&=& \dfrac{6 C_1 C_2 \kappa r^3}{\rho_0}(5\cos^2\theta-1), \nn \\
W(r,\theta)&=&w_0+ \nn \\
        &&\dfrac{(w_0-1)r^2}{88}(23 C_3+11 \rho_0+20 C_3\cos(2\theta)), \nn \\
A_\theta(r,\theta)&=& C_2 \kappa r^3 \sin\theta, \nn \\
A_\varphi(r,\theta)&=&C_3 r^2 \sin^2\theta, \nn \\
B_\theta(r,\theta)&=&C_2 r^3 \sin\theta,\nn \\
B_\varphi(r,\theta)&=&-C_3 r^2 \sin^2\theta,\nn \\
\rho(r,\theta)&=&\rho_0+r^2 (C_4+\dfrac{6 C_2 \kappa}{\rho_0}+3 C_4 \cos(2\theta)),
\label{typeIsol}
\eea
where $C_i,~ w_0,\rho_0$ are integration constants. We keep only those
independent integration constants for which the solution
satisfies the finite energy conditions
and the symmetry under reflection $z\rightarrow -z$.
A local finite energy solution in the asymptotic region at space infinity
is given by the following expressions
%{\scriptsize {\bf A82J8July5solGiFi=1ViThetaBCHCK.nb}}
%in this file the lambda equals 3 k^2/2
\bea
f(r,\theta)&=&1-2 \tC_1 \exp(-r), \nn  \\
G(r,\theta)&=&1+\dfrac{\tC_2\kappa}{r^2} \exp(-\varkappa r), \nn \\
W(r,\theta)&=&1+\dfrac{\tC_3(1+r)}{r} \exp(-r)  , \nn \\
A_\theta(r,\theta)&=&-\tC_2 \kappa \exp(-\varkappa r)\sin\theta , \nn \\
A_\varphi(r,\theta)&=& 1-\cos(2\theta)-\dfrac{\tC_4}{r}+\tC_5 \kappa \exp(-\varkappa r)
                         \sin^2\theta   , \nn \\
B_\theta(r,\theta)&=& -\tC_2\exp(-\varkappa r) \sin\theta  ,\nn \\
B_\varphi(r,\theta)&=& \dfrac{\tC_4}{r}+\tC_5\exp(-\varkappa r)\sin^2\theta ,\nn \\
\rho(r,\theta)&=&1+\dfrac{\tC_6}{r}\exp(-k r),\nn \\
\varkappa^2&\equiv& 1+\frac{1}{\kappa}, \label{solinf1}
\eea
where we keep only leading terms.
Note that the solutions for azimuthal potentials
$A_\varphi,~B_\varphi$ include the term $\dfrac{\tC_4}{r}$ which provides
long range behavior of the electromagnetic potential and corresponds to
the dipole magnetic moment of the sphaleron solution.
We will neglect this term in setting boundary conditions
since we are interested in
the magnetic solution with only one, azimuthal, non-vanishing magnetic field component.
This can be reached by setting the constraints
\bea
A_\varphi+\tC_\varphi=0,~~~~~~~B_\varphi=0. \label{constr3}
\eea
Notice, the above constraint has exactly the same structure
as the gauge invariant quantity in (\ref{calA}).
An additional variational analysis with a non-vanishing
trial function for $\tC_\varphi$ shows that minimum of
total energy is reached precisely when the constraint
(\ref{constr3}) is fulfilled.

The structure of the boundary conditions provides a minimal
topological magnetic flux $2 \pi$ of the azimuthal magnetic field
through the half plane $P: \{y=0,~x\geq0\}$
\bea
\Phi_\varphi&=& \int dr d\theta {\cal F}_{r\theta}=\int dr d\theta H_{r\theta}=2 \pi.
\eea
In lowest order approximation we choose a simple radial dependence
for the trial functions except for the
function $A_\theta (r,\theta)$. The gauge field $A_\theta (r,\theta)$
provides a dominant contribution to the azimuthal magnetic field,
so we keep two leading order terms in Fourier series expansion for that function
to ensure a non-vanishing value of the first mode $A^{(1)}_\theta(r)$
\bea
A_\theta(r,\theta)&=&A^{(1)}_\theta(r) \sin \theta+A^{(2)}_\theta(r) \cos(2 \theta).\label{Atheta}
\eea
With this setup we minimize the energy functional (\ref{energy}).
The obtained variational profiles for the trial functions are shown in Fig. 3.
The energy density in cylindrical coordinates has two
maximums shown in Fig. 4, which correspond to two parallel
circles coincided with the center lines of two tori in Cartesian coordinates,
Fig. 5a. The total energy estimate is $1.36$ TeV.
The surfaces of constant values for the energy density form tori
or discs according to energy density values, Figs. 5a, 5b.
%{\scriptsize {\bf Ap11EnergyfQ-5hhPLOTsimp.nb, Ap10M16EnergyfQ-5hhPLOTS.nb, Ap7M16EnergyfQ-5.nb}}
\begin{figure}[htp]
\centering
\includegraphics[width=65mm,height=45mm]{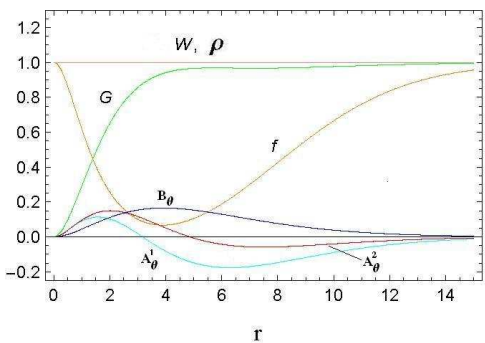}
\caption[fig3]{Trial functions: $f$-orange, $G$-green, $W$ - dashed, $A^{(1)}_\theta$ - cyan,
$A_\theta^{(2)}$ - red, $B_\theta$ - blue, $\rho$ - pink}\label{Fig3}
\end{figure}
\begin{figure}[htp]
\centering
\includegraphics[width=65mm,height=45mm]{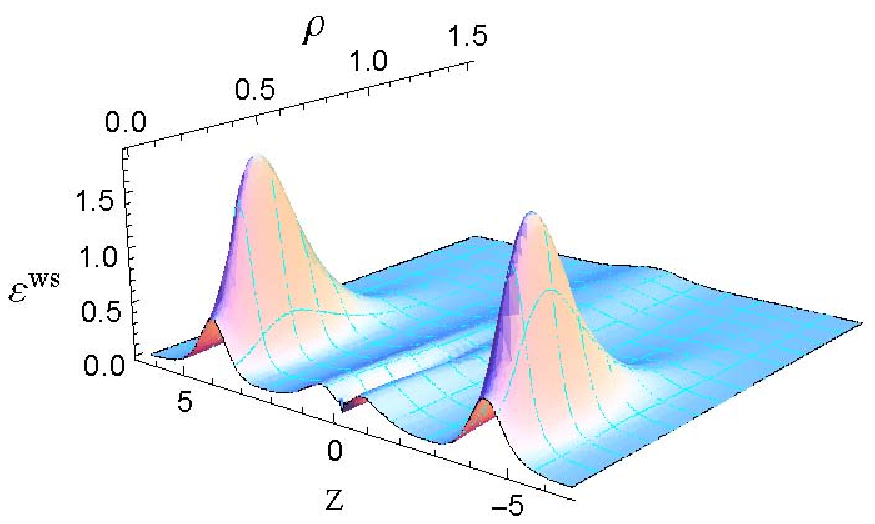}
\caption[fig4]{Energy density in cylindrical coordinates $(\rho, z)$. }\label{Fig4}
\end{figure}
\begin{figure}[htp]
\centering
\subfigure[~]{\includegraphics[width=40mm,height=40mm]{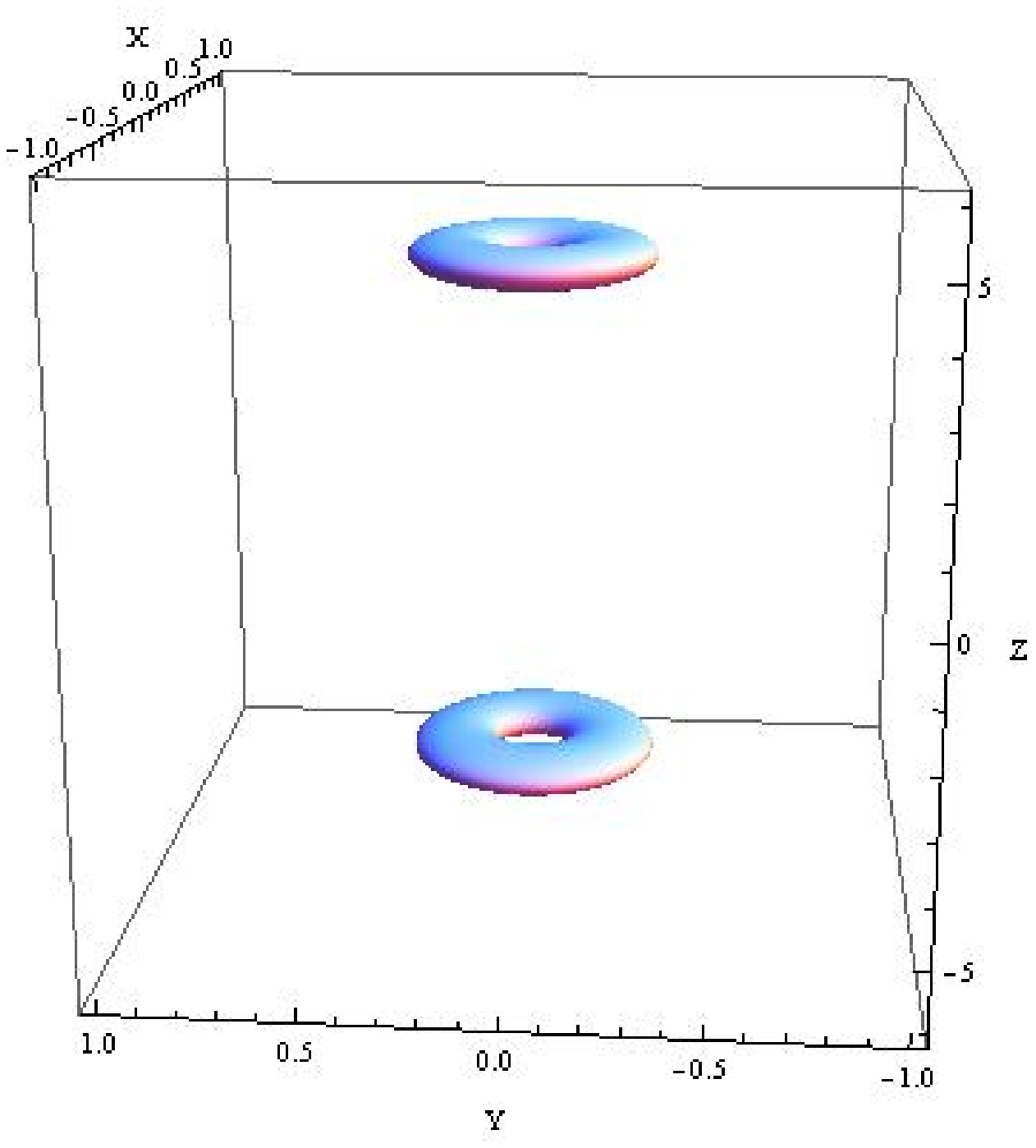}}
%\caption[AFZplot11]{$m=1,\,n=1$}
\hfill
%A22A16fQF0=1-1PLOTS.nb, or its reduced version A22A16PLOTS.nb
\subfigure[~]{\includegraphics[width=40mm,height=40mm]{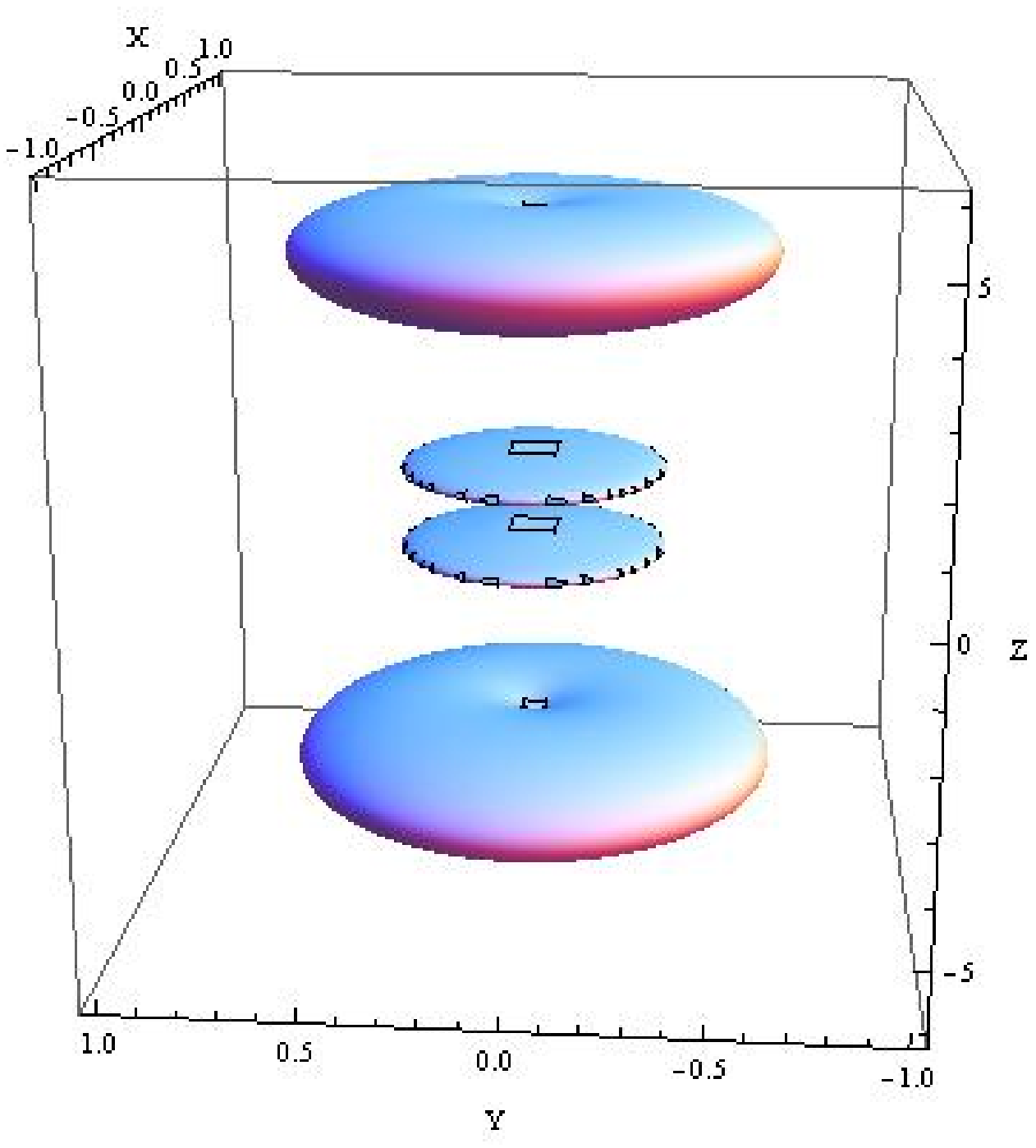}}
\caption[ fig5]{Three-dimensional contour plots for the energy density
surfaces in Cartesian coordinates $(X,Y,Z)$ for selected values:
${\cal E}^{WS}=1.5$, (a), and
${\cal E}^{WS}=0.157$, (b).}\label{Fig5}
\end{figure}
%Variational profiles for all
%trial functions are regular.
Low resolution leading to
seemed irregularities in Fig. 5b (depicted by small squares)
is caused by insufficient computer memory.
% in constructing three-dimensional contour plot.
%A total energy of such field configuration is $1.36$ TeV,

%\vspace*{2mm}
\subsection{ Type II boundary conditions}
%\vspace*{2mm}

The local solution to the Weinberg-Salam equations
near the origin $r=0$ (\ref{typeIsol})
is not unique. Instead of Taylor series expansion
for the trial functions one can apply
perturbation theory to construct approximate solutions
to the equations of motion.
One can introduce a small parameter $\alpha$
which can be associated with the coupling constant
$g$ or $g'$ assuming they are small parameters.
The perturbation theory with the parameter $\alpha$
can be consistently constructed
since we use the dimensionless variables,
so the couplings $g,~g'$ are absorbed in
dimensionless field variables as in (\ref{energy}).
With this a new local solution near the origin
can be obtained in the first order of the perturbation theory
%{\scriptsize {\bf 1M8Jul13G0=0CHECKSOL.nb,
%  C9 ->C6, C21->C3, C4->C6, C6->C4,C10->C5
% A7M10M8Jul13G0=0CHECKSOLcaseF=0.nb}}
\bea
f(r,\theta)&=&f_0+\alpha C_1 (3-f_0) r, \nn \\
%-\dfrac{F_0}{12 C_1 \rho_0} r^2\cdot \nn \\
%             && \big(-8 C_22 F_0+2 C_1 C_3 (2 F_0-1) \rho_0+C_4 \cos(2 \theta)\big )\Big) , \nn  \\
G(r,\theta)&=&G_0+\alpha C_1 r , \nn \\
%+ \nn \\   &&\dfrac{1}{20} r^3 (-5 C_2 + C_1 \rho_0^2 + 25 C_2 \cos(2 \theta)), \nn \\
W(r,\theta)&=&w_0+\alpha C_2 r^4 \cos^2 \theta, \nn \\
A_\theta(r,\theta)&=&\alpha \dfrac{C_1(3-f_0)\rho_0^2}{12} r^3 \sin\theta , \nn \\
A_\varphi(r,\theta)&=&\alpha C_3 r^2 \sin^2\theta, \nn \\
B_\theta(r,\theta)&=&-\alpha \dfrac{C_1 (3-f_0)\rho_0^2}{12} r^3 \sin\theta,\nn \\
B_\varphi(r,\theta)&=& \alpha C_4 r^2 \sin^2 \theta,\nn \\
\rho(r,\theta)&=&\rho_0+\alpha C_5 r^2 \big (1+3 \cos(2 \theta)\big ),
\eea
%\bea
%f(r,\theta)&=&f_0+g\Big( C_1 (3-f_0) r +O(r^2) \Big), \nn \\
%%-\dfrac{F_0}{12 C_1 \rho_0} r^2\cdot \nn \\
%%             && \big(-8 C_22 F_0+2 C_1 C_3 (2 F_0-1) \rho_0+C_4 \cos(2 \theta)\big )\Big) , \nn  \\
%G(r,\theta)&=&G_0+g \big (C_1 r+ \nn \\
%      &&\dfrac{1}{20} r^3 (-5 C_2 + C_1 \rho_0^2 + 25 C_2 \cos(2 \theta))\big), \nn \\
%W(r,\theta)&=&w_0+g C_3 r^4 \cos^2 \theta, \nn \\
%A_\theta(r,\theta)&=& g \Big(\dfrac{C_1(3-f_0)\rho_0^2}{12} r^3 \sin\theta+O(r^5)\Big) , \nn \\
%A_\varphi(r,\theta)&=&g \Big(C_4 r^2 \sin^2\theta +O(r^4) \Big), \nn \\
%B_\theta(r,\theta)&=&g \Big (-\dfrac{C_1 (3-f_0)\rho_0^2}{12} r^3 \sin\theta +O(r^5) \Big),\nn \\
%B_\varphi(r,\theta)&=& g \Big (C_5 r^2 \sin^2 \theta+O(r^4) \Big),\nn \\
%\rho(r,\theta)&=&\rho_0+g \Big (C_6 r^2 \big (1+3 \cos(2 \theta)\big )+O(r^4)\Big ),
%\eea
where $C_i,~f_0,G_0,w_0,\rho_0$ are integration constants. Note that,
due to the highly non-linear structure of the equations of motion
to obtain the solution for $f,W$ in first order approximation
one has to obtain solutions for other functions up to the second
order of the perturbation theory. We present in the solution only
terms of first order in $\alpha$ and leading terms in Taylor series expansion.
We retain only those
integration constants which provide the finite energy condition and
symmetry under reflection
$z\rightarrow -z$. In addition, we set $G_0=0$ whereas
$f_0,w_0,\rho_0$ are treated as free variational parameters.

A local finite energy solution to Weinberg-Salam equations in the
asymptotic region $r\simeq \infty$
is given by the same equations  (\ref{solinf1}) as in the case of type I
boundary conditions.
%{\scriptsize {\bf A82J8July5solGiFi=1ViThetaBCHCK.nb}}
%in this file the lambda equals 3 k^2/2
%\bea
%F(r,\theta)&=&1+\tC_1 \exp(-r), \nn  \\
%G(r,\theta)&=&1+\dfrac{\tC_2\kappa}{r^2} \exp(-\varkappa r), \nn \\
%W(r,\theta)&=&1+\dfrac{\tC_3(1+r)}{r} \exp(-r)  , \nn \\
%A_\theta(r,\theta)&=&-\tC_2 \kappa \exp(-\varkappa r)\sin\theta , \nn \\
%A_\varphi(r,\theta)&=& 1-\cos(2\theta)-\dfrac{\tC_4}{r}+\tC_5 \kappa \exp(-\varkappa r)
%                         \sin^2\theta   , \nn \\
%B_\theta(r,\theta)&=& -\tC_2\exp(-\varkappa r) \sin\theta  ,\nn \\
%B_\varphi(r,\theta)&=& \dfrac{\tC_4}{r}+\tC_5\exp(-\varkappa r)\sin^2\theta ,\nn \\
%\rho(r,\theta)&=&1+\dfrac{\tC_6}{r}\exp(-k r),~~~~~\varkappa^2\equiv 1+\frac{1}{\kappa},
%\eea
%where we keep only leading terms in degrees of $O\big(\dfrac{1}{r}\big)$.
As in the previous section we impose constraints (\ref{constr3})
implying only azimuthal non-vanishing magnetic field component.
The corresponding total magnetic flux around the $Z$-axis is $2\pi$.
We keep two leading order terms in Fourier series expansion for
the function $A_\theta$ as in (\ref{Atheta}).

The results of the variational procedure
of minimizing the energy functional for the trial functions
are presented in Fig. 6.
%{\scriptsize {\bf Ap11EnergyfQ-5hhPLOTsimp.nb, Ap10M16EnergyfQ-5hhPLOTS.nb, Ap7M16EnergyfQ-5.nb}}
\begin{figure}[htp]
\centering
\includegraphics[width=65mm,height=50mm]{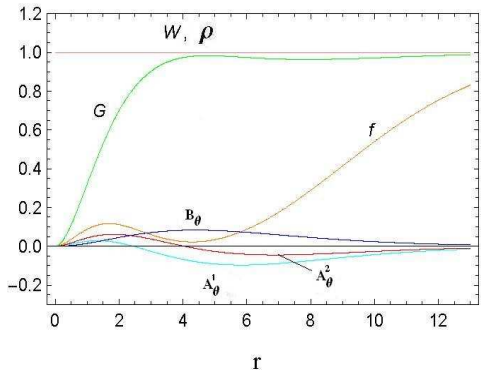}
\caption[fig6]{Trial functions: $f$-orange, $G$-green, $W$ - dashed, $A^{(1)}_\theta$ - cyan,
$A_\theta^{(2)}$ - red, $B_\theta$ - blue, $\rho$ - pink}\label{Fig6}
\end{figure}
%A27A16fQF0=f0-1hh.nb
\begin{figure}[htp]
\centering
\includegraphics[width=65mm,height=45mm]{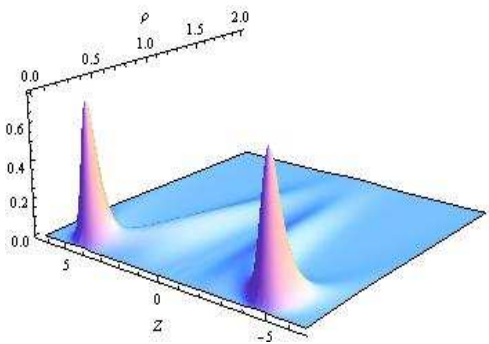}
\caption[fig7]{Energy density in cylindrical coordinates $(\rho, z)$.}\label{Fig7}
\end{figure}

Notice that the minimization procedure induces a
zero value for the parameter $f_0$. We provide
two types of boundary values for the function $f(r,\theta)$
since a solution with type I boundary conditions
may exist even though it might be rather unstable.
The energy density profile in cylindrical coordinates is plotted in
Fig. 7.
Qualitatively, the picture for the energy density is similar to the one in the case
of a magnetic field configuration corresponding to type I boundary conditions:
two maximums of the energy density are located along two circles with centers
on the $Z$-axis. A total energy is $0.98$ TeV, which is much
less than the energy value $18$ TeV of the
magnetic field solution in the $CP^1$ model.
Let us consider the electric current $\hat J_n$
which is responsible for the gauge invariant Abelian
magnetic field ${\cal F}_{mn}$
\bea
\hat J_n=-\partial^m {\cal F}_{mn}.
\eea
In Fig. 8 we show that
vector lines of the electric current $\hat J_n$
have a regular structure which has been checked at various scales
in the interval $r=(0.01, 30)$.
%May10A27STREAMPLOTS.nb
\begin{figure}[htp]
\centering
\includegraphics[width=65mm,height=45mm]{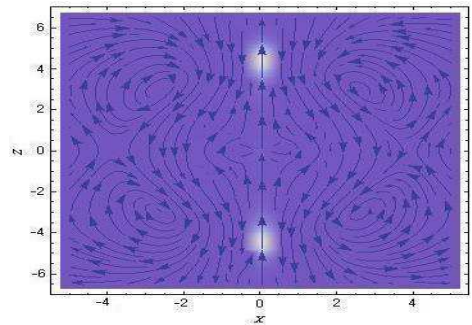}
\caption[fig8]{Vector stream plot for the electric current $\hat J_n$
in the plane $(X,Z)$
 in Cartesian coordinates.}\label{Fig8}
\end{figure}

Certainly, the energy minimization
procedure based on a restricted variational
ansatz does not guarantee that
local solutions near the origin and at infinity will match
in the whole space. A rigorous approach should
include Fourier series expansion for
all fifteen variational
functions within the most general axially-symmetric ansatz,
or one should solve a complicated system of PDEs
which has not been done so far except for cases of
known solutions described by DHN ansatz.
% We hope that problem
%of exact numeric solving
%equations of motion for axially-symmetric magnetic solutions
%admitting helical structure will be resolved in further studies.

In conclusion,
we have demonstrated that interaction structure of the
gauge and Higgs bosons implies existence of magnetic field
configurations with energy upper bound near 1 TeV
which is essentially less than energy of monopole like solutions in
the Weinberg-Salam model. This would give rise to an
attractive possibility for search of respective
new bound states of $W,Z$ and Higgs bosons
in concurrent experimental facilities.

\acknowledgments
%{\bf Acknowledgements}

Authors thank E. Tsoy for numerous useful discussions and
L. Graham for careful reading our paper.
The work is supported by NSFC (Grants 11035006 and 11175215),
the Chinese Academy of Sciences visiting professorship for senior international
scientists (Grant No. 2011T1J3), and by UzFFR (Grant F2-FA-F116).
%\clearpage
%\vspace{2mm}

\end{document}